\input harvmac
%\draftmode
\def \la{\longrightarrow}

 \def \upa {\uparrow}
 \def \nea {\nearrow}

\def \ha{{\textstyle{1\over 2}}}

\def \a {\alpha}
\def \b {\beta}
\def\mapa#1{\smash{\mathop{\longrightarrow }\limits_{#1} }}

\def \s {\sigma}
\def \p {\phi}

\def \n {\nu}

\def \t {\theta}
\def \td {\tilde }

\def \inv {^{-1}}
\def \ov {\over }

\def\ww {\omega _{km } }

\def\mn { {(-k,-m)} }

\def\ww {\omega _{km } }

\def\rss { ${\bf R}^{9}\times S^1 \times S^1 $}

\def \lr { \lref}
\def\np {{  Nucl. Phys. }}
\def \pl {{  Phys. Lett. }}
\def \mpl {{ Mod. Phys. Lett. }}
\def \prl {{  Phys. Rev. Lett. }}
\def \pr  {{ Phys. Rev. }}

\baselineskip8pt
\Title{
\vbox
{\baselineskip 6pt{\hbox
{Imperial/TP/96-97/20 }}{\hbox{hep-th/9701188}} {\hbox{
  }}} }
{\vbox{\centerline {T-duality in M-theory and supermembranes}
%oscillating states  in M-theory}
\vskip4pt
%\centerline { }
%\vskip4pt
% \centerline { }
}}
\vskip -27 true pt
\centerline  {  J.G. Russo{\footnote {$^*$} {e-mail address:
jrusso@ic.ac.uk. } }}
\smallskip \medskip
\centerline{\it Blackett Laboratory,  Imperial College,  London  SW7 2BZ}
%\smallskip
%\centerline{\it    }
\bigskip
\medskip
\centerline {\bf Abstract}
\medskip
\baselineskip10pt
\noindent
The $(q_1,q_2)$ $SL(2,{\bf Z})$ string bound states of type IIB superstring theory
admit
two inequivalent (T-dual) representations in eleven dimensions
in terms of a fundamental 2-brane.
In both cases, the spectrum of membrane oscillations can be determined
exactly in the limit $g^2\to \infty $, where $g^2$ is the type IIA
string coupling.
We find that the BPS mass formulas agree, and
reproduce the BPS mass spectrum  of the $(q_1,q_2)$ string bound state.
In the non-BPS sector, the respective mass formulas
apply in different  corners of the moduli space. 
The axiomatic requirement of T-duality in M-theory permits to derive
a discrete mass spectrum in a (thin torus)
region where standard supermembrane theory undergoes instabilities.

\medskip
%%%%%%%%%%%%%%%%%%%%%%%%%%%%%%%%%%%%%%%%%%%%%%%%%%%%%%%%%
\Date {January 1997}
%%%%%%%%%%%%%%%%%%%%%%%%%%%%%%%%%%%%%%%%%%%%%%%%%%%%%%%%%%%%%%%%%%%
\noblackbox
\baselineskip 14pt plus 2pt minus 2pt
%\baselineskip 20pt plus 2pt minus 2pt
%%%%%%%%%%%%%%%%%%%%%%%%%%%%%%%%%%%%%%%%%%%

\lr \dus { M.J. Duff and  K.S. Stelle, \pl B253 (1991) 113.}

\lr\hos{G.T.~Horowitz and A.~Strominger, Nucl. Phys. { B360}
(1991) 197.}
\lr \duf { M.J. Duff, P.S. Howe, T. Inami and K.S. Stelle, 
\pl B191 (1987) 70. }

\lr\dvv{R. Dijkgraaf, E. Verlinde and H. Verlinde, hep-th/9603126.}
\lr\gibb{G.W. Gibbons and P.K. Townsend, \prl  71
(1993) 3754, hep-th/9307049.}

\lr \ttt{P.K. Townsend, hep-th/9512062.}
\lr \papd{G. Papadopoulos and P.K. Townsend, \pl B380 (1996) 273, hep-th/9603087.}

\lr\dulu{ M.J. Duff and J.X. Lu, \np B347 (1990) 394;
E.~Sezgin and R.~Percacci, \mpl A10 (1995) 441.}
\lr \duflu { M.J. Duff and J.X. Lu, \np B354 (1991) 141. } 
\lr \pol { J. Polchinski, \prl 75 (1995) 4724,  hep-th/9510017.} 
\lr \iz { J.M. Izquierdo, N.D. Lambert, G. Papadopoulos and 
P.K. Townsend,  \np B460 (1996) 560, hep-th/9508177. }

\lr \US{M. Cveti\v c and  A.A.  Tseytlin, 
\pl {B366} (1996) 95, hep-th/9510097;   hep-th/9512031.  
}

\lr \green{M.B. Green and M. Gutperle, hep-th/9604091.}

\lr \berg{E. Bergshoeff, C. Hull and T. Ort\' \i n, \np B451 (1995) 
547.}

\lr \gibbon{G.W. Gibbons, \np B207 (1982) 337. }
\lr \hullo{C.M. Hull, \pl B139 (1984) 39. }

\lr \tset  {A.A.  Tseytlin,  \np B475 (1996) 179, hep-th/9604035.}
\lr \john {J.H.  Schwarz, \pl B360 (1995) 13 (E: B364 (1995) 252).}
\lr \johnt {J.H.  Schwarz, \pl B367 (1996) 97, 
 hep-th/9510086. }

\lr \townelev{ P.K. Townsend, \pl B350 (1995) 184, hep-th/9501068.  }
\lr \klts{I.R. Klebanov and A.A. Tseytlin,
 \np B475 (1996) 179,
hep-th/9604166. }
\lr \cvets{ M. Cveti\v c  and A.A. Tseytlin, \np B478 (1996) 181, 
hep-th/9606033. }
 
\lr\paptkk{
G. Papadopoulos  and P.K. Townsend, hep-th/9609095. }

\lr\papadop{
G. Papadopoulos, hep-th/9604068. }

\lr \witten {E. Witten, \np B460 (1995) 335.}

\lr \tow {P.K. Townsend,  hep-th/9609217. } 

\lr \schwa  {J.H. Schwarz,
% Lectures given at Spring School on String Theory, Gauge
%Theory and Quantum Gravity, Trieste, Italy, 18-29 Mar 1996,
hep-th/9607201.  }
\lr \duff { M.J. Duff, hep-th/9608117. }
\lr\dufmb{ M.J. Duff,  hep-th/9611203.}
\lr \bergsh{ E. Bergshoeff, E.  Sezgin and P.K. Townsend, \pl B189 (1987)
75.}
\lr \doug {M.R. Douglas,  hep-th/9512077.}
\lr \gaunt {J. Gauntlett, D. Kastor and J. Traschen, hep-th/9604189.}
\lr \aspin {P. Aspinwall,  hep-th/9508154.   }
\lr \lifsh {G. Lifschytz, hep-th/9610125. }

\lr\grepap{
M.B. Green, N.D. Lambert, G. Papadopoulos  and P.K. Townsend, 
hep-th/9605146. }

\lr\banks{
T. Banks, W. Fischler, S.H. Shenker and L. Susskind, 
   hep-th/9610043; M.~Berkooz and M.R.~Douglas, hep-th/9610236;
O.~Aharony and M.~Berkooz, hep-th/9611215;
T.~Banks, N.~Seiberg and S.~Shenker, hep-th/9612157.}

\lr \tser{A.A. Tseytlin, hep-th/9609212. }

\lr\costa { M.S. Costa, hep-th/9609181.}

\lr \schmid {C.  Schmidhuber, \np B467 (1996) 146,   hep-th/9601003.  }
\lr \tsetli{ A.A. Tseytlin, \np B469 (1996) 51, hep-th/9602064.  } 
\lr\russo {J.G. Russo, 
%``Supermembrane dynamics from multiple interacting strings",
hep-th/9610018 .}
\lr\dewitt { B. de Wit, J. Hoppe and H. Nicolai, 
\np  {B305} [FS 23] (1988) 545.}
\lr\dufi{M.J. Duff, T. Inami, C. Pope, E. Sezgin and K. Stelle,
\np {B297} (1988) 515.}
\lr\bst{E. Bergshoeff,  E. Sezgin and Y. Tanii, \np {B298}
(1988) 187.}
\lr \bergsh { E. Bergshoeff,  E. Sezgin and P.K. Townsend,
%\pl { B189} (1987) 75;
 Ann. Phys. {185} (1988) 330.} 
\lr\dewit { B. de Wit, M. L\" uscher and H. Nicolai, 
\np  {B320} (1989) 135.}

\lr\dddo{M.R. Douglas, D. Kabat, P. Pouliot and  S.H. Shenker,
  hep-th/9608024. }
\lr\beh {K. Behrndt, E. Bergshoeff and B. Janssen, hep-th/9604168
(revised).}
\lr\verl{E. Verlinde, \np B445 (1995) 211.}

\lr\sen{A. Sen, \pr D53 (1996) 2964, 
hep-th/9510229;   \pr D53 (1996)  2874,  hep-th/9511026.}

\lr\reviu {For recent reviews, see:
J.H. Schwarz, hep-th/9607201; M.J. Duff, hep-th/9608117;
J.~Polchinski, hep-th/9611050; P.K. Townsend, hep-th/9612121.}

\lr\kaa{ E.A. Bergshoeff, R. Kallosh and T. Ort\'\i n, 
    \pr D47 (1993) 5444.}

\lr\rutse {J.G. Russo and A.A. Tseytlin, hep-th/9611047.}

\lr\myers {J.C. Breckenridge, G. Michaud and R.C. Myers,
hep-th/9611174.}
\lr\papa  {M.S. Costa and G. Papadopoulos, hep-th/9612204.}
\lr\wtayl  {W. Taylor, hep-th/9611042.}
\lr\sussk  {L. Susskind, hep-th/9611164;
O.J. Ganor, S.~Ramgoolam and W.~Taylor,
hep-th/9611202. }

%%%%%%%%%%%%%%%%%%%%%%%
%\newsec{The dual eleven dimensional backgrounds}
%%%%%%%%%%%%%%%%%%%%%%%%

Recently, the eleven dimensional description of different
BPS bound states of type~II superstring theories was worked out \rutse .
The determination  of the 2-brane oscillation
spectrum in a certain limit,
together with the existence of 11d 2-brane backgrounds connected by T-duality, will be exploited here to perform a concrete test of
T-duality in M-theory\foot{
By which here  we  mean the strong coupling limit of the type IIA superstring
(for recent reviews, see \reviu ).
A search for duality symmetries using 3d sigma-model description
was undertaken in \dulu .
More recent discussions on T-duality using D-brane lagrangians 
and (M)atrix theory \banks\ are in ref. \refs {\wtayl, \sussk }.} 
in the BPS sector, and to evade supermembranes instabilities \dewit\ 
by going to the dual representation.

In standard perturbative string theory, T-duality symmetry
is the assertion  that  conformal field theories corresponding to
two backgrounds related by a T-duality transformation are equivalent.
In particular, the one-loop partition function is the same 
for both systems, and 
there is a one-to-one correspondence between the spectra.
Related to these properties is the fact that the effective
action is T-dual invariant to all orders in the $\a' $ expansion.
%In string perturbation theory,
%two  backgrounds connected by a T-duality transformation
%describe equivalent physics.
In eleven dimensional supergravity, solutions with  two or more isometries
are also related by similar transformations \berg .
Whether this property is to hold beyond leading order in M-theory
remains to be proved.
In particular, it strongly relies on a complete
matching of the mass spectra of excitations of the dual backgrounds.
This is the case in string theory, where 
winding states are crucial in order
for T-duality to be an exact  symmetry to all orders in perturbation theory.

The example considered here will be based on the 
$(q_1,q_2)$ 1/4 supersymmetric string bound states with a momentum boost 
$w_0$ along the string \refs {\john , \witten}, with an extra isometry along $y_3$. 
In the string frame, the corresponding classical solution 
is given by ($i=1,...,7$)
\eqn\uno{
ds^2_{10 B} =   K^{1/2} \left( \td H_2^{-1} \left[ -dt^2 + dy^2_1 +  
 (H_2-1) (d t- d y_1)^2 \right] + dy_3^2+dx_i dx_i \right)\ ,
}
$$ e^{2\p}=   \td H_2\inv  K^2  \ , \ \ \ \  \ \   \chi =  \sin \t  
 \cos \t\  W K^{-1}  \ ,   
$$ 
$$
B^{(1)}_{ty_1}  =  - \cos \t\ W  \td H_2^{-1}  \ , \ \  \ \ \ \ 
B^{(2)}_{t y_1}  =  - \sin \t \ W  \td H_2^{-1} \ ,  
$$
with 
$$
H_2=1 + {Q\ov r^5 }\ ,\ \  \td H_2=1+{\td Q\ov r^5 }\ ,\ \ 
\ K=1+\sin^2 \theta \  {\td Q\ov r^5 }\ ,
\ \ \ W= {\td Q\ov r^5}\ ,\ \ \ r^2=x_ix_i\ ,
$$ 
$$
Q=c_0 {w_0 \ov R_1' }\ ,\ \ \ \ \td Q=c_0 {nR_1'\over\a'}
\sqrt{q_1^2+q_2^2\tau_2^2}\ ,\ \ \ \ c_0={\kappa_9^2\ov 3 \omega _7}\ ,
$$
and
$$ 
\cos\theta = {q_1\ov \sqrt{ q_1^2 + q_2^2\tau_2^2}}\ ,\ \ \ \
\sin\theta = {q_2 \tau_2\ov \sqrt{ q_1^2 + q_2^2\tau_2^2}}\ ,\ \ \ \ \ \tau_2=e^{-\phi_0}={R_1\ov R_2}\ ,\ \ \ \ R_1\equiv {\a'\ov R_1'}\ .
$$
Here the $S^1$-coordinate $y_1$ has period $2\pi R_1'$, whereas $R_2$
represents the radius of the ``eleventh" dimension $y_2$ that will
appear in a moment. $R_1$ is the radius of the type IIA 
$y_1$-coordinate.
The special cases are the NS-NS string 
($q_1=1, q_2=0$, i.e. $\t=0, \  K=1$), and the R-R string 
  ($q_1=0, q_2=1$, i.e.
  $\t={\pi \ov 2} ,   \  K=\td H_2 $). 
 $T$-duality  along $y_1$ transforms  it into a 
$D=10$ type IIA background, which  can be 
interpreted as a 1/4 supersymmetric non-threshold
bound state of a fundamental string and a D0-brane boosted along the string
direction, $ 1_{NS} + 0_R +\upa $
(more solutions representing non-threshold bound states of type IIA string theory were obtained in refs. \refs{\rutse, \myers, \papa}),
\eqn\dos{
ds^2_{10A}=H_2^{-1} K^{1/2} \big[ -dt^2 +dy_1^2+
W K^{-1}(dt-\cos\theta \ dy_1)^2\big]
+K^{1/2} \big( dy_3^2+ dx_idx_i \big)\ ,
}
$$
B_{ty_1}=H_2^{-1}\ ,\ \ \ e^{2\phi }=H_2^{-1} K ^{3/2}\ ,
\ \ \ A=\sin\theta \ W  K^{-1} (dt-\cos\theta \ dy_1)\ .
$$
The lift to $D=11$ gives the solution found in \rutse \
\eqn\tres{
{\rm (A):}\ 
\ \ \ ds^2_{11} = H^{-2/3}_2   [-    dt^2  + dy^2_1 +  dy^2_2 + 
(\td H_2-1) (dt - dz_1)^2 ] +  
H_2^{1/3} \big( dy_3^2+ dx_i dx_i \big)  , 
}
$$  
C_3  =  H_2\inv dt\wedge  dy_1\wedge d y_2 \ ,
\ \ \ \ z_1=\cos \t\ y_1 + \sin \t\ y_2 \ ,
$$ 
which represents a 2-brane wrapped around the torus $(y_1,y_2)$
 with winding $w_0$
and  quantized momentum with components 
$(l_1/R_1, l_2/R_2)=(nq_1/R_1,nq_2/R_2)$, and  modulus ${n\ov R_1} 
\sqrt {q_1^2 + q_2^2\tau_2^2}$.
The total momentum  is proportional to
  the tension of the $(q_1,q_2)$ string.
Above  we have considered  the case of the  rectangular 2-torus
  $\tau=i\tau_2$, or the  vacuum
  $ \rho_0 = \chi_0 + i e^{-\p_0} =ie^{-\p_0}$  .
   Generalization  to the case of an arbitrary 
   2-torus is straightforward.
    
%%%%%%%%%%%%%%%%%%%%%%%%%%%%%%%%%%%%%%%
%\subsec { Another representation}
%%%%%%%%%%%%%%%%%%%%%%%%%%%%%%%%%%%%%%%
Starting with the same $(q_1,q_2)$ $SL(2,{\bf Z})$ string 
bound state, we now perform
a T-duality transformation along a direction perpendicular
to the string, $y_3$. In this process, the D-string becomes a D2-brane.
We then lift the resulting solution to eleven dimensions, obtaining
\eqn\dualo{
{\rm (B):}\ \ \ \ ds^2_{11}=\td H_2^{-2/3} \big[ -dt^2+dy_1^2+d z_2^2  + (H_2-1)(dt-dy_1)^2 \big] + \td H_2^{1/3}(dz_3^2+dx_idx_i)\ ,
}
$$  
C_3  =  \td H_2\inv dt\wedge  dy_1\wedge d z_2 \ ,
$$ 
$$
z_2=y_2 \cos \theta + y_3\sin\theta\ ,\ \ 
z_3=-y_2 \sin \theta +y_3 \cos\theta\ .
$$
This is a 2-brane with a momentum boost $w_0/R_1'$ along $y_1$, with
one leg on $y_1$ and the other wrapped around the $(q_1,q_2)$ cycle
of the torus $(y_2,y_3)$.
When $w_0=0$ (corresponding to the IIB string bound state with
zero momentum), this becomes a {\it static } background, whereas
in the representation (A) the $w_0=0$ case has non-zero momentum. 
There exists no reparametrization
which connects both backgrounds;  
from the viewpoint of supergravity effective field theory
the two solutions are inequivalent. 
They are, however, of the same form, where the roles of $Q$ and $\td Q $
(i.e. winding number and total momentum) are interchanged, but in addition
the membrane  is wrapped in a different way around the 3-torus.

Schematically, the two backgrounds are obtained by the 
sequences
$$
{\rm (A):}\ \ \ \ \ \ \ \ \ \ \ 1_{NS}+1_R+\upa\ \mapa{T_{y_1}}\ \upa +0_{R}+ 1_{NS}\ 
\mapa {\rm lift} \ 2+\nea
$$
$$
{\rm (B):}\ \ \ \ \ \ \ \ \ \ \ 1_{NS}+1_R+\upa\ \mapa{T_{y_3}}\ 1_{NS} +2_{R}+ \upa \ 
\mapa {\rm lift} \ 2+\upa\ 
$$
From the type IIA perspective, in going from one representation to another, 
one has the  chain
\eqn\seis{
{\rm (A)}\ =\ \upa +0_{R}+ 1_{NS}\  \mapa{T_{y_1}} \ 1_{NS}+1_R+\upa\  
\mapa{T_{y_3}}\ 1_{NS} +2_{R}+ \upa \ =\ {\rm (B)}
}
\def\n{ {\bf n} }

%%%%%%%%%%%%%%%%%%%%%%%%%%%%%%%%%%%%%%%%%%%%%%%%%%%%%%%
%\newsec{The mass spectrum for the representation (A) }
%%%%%%%%%%%%%%%%%%%%%%%%%%%%%%%%%%%%%%%%%%%%%%%%%%%%%%%

The zero-mode part of the membrane spectrum for the initial 
system (A) was investigated in ref.~\john , and
the oscillator part was calculated in  ref. \rutse , 
where the BPS sector of the spectrum was also identified.\foot{ 
For previous discussions of the membrane spectrum on
\rss  , see \refs {\bergsh  \dufi -\russo }; for a  recent review
on supermembranes, see \dufmb\ and references there.
}
The mass formula 
\eqn\nombre{
M^2_{\rm (A)}=M_0^2 + {2\ov \a'} { H_{\rm (A)}}\ ,\ 
}
\eqn\alfa{
 (\a')^{-1}\equiv 4\pi ^2 R_2 T_3\ ,
}
contains the  zero-mode part $M_0^2$,
which can be calculated exactly for all $g^2=R_2^2/\a '$, 
and it is given by
\eqn\mmzer{
M^2_0={l_1^2\ov R_1^2} + {l_2^2\ov R_2^2} + {w_0^2 R_1^2\ov 
\a  ^{\prime 2}}\ ,
}
and an oscillator part ${ H}_{\rm (A)}$, which can be determined exactly in the limit
 $g^2\to \infty $, with $\a ' $ and $\tau_2=R_1/R_2$ fixed
(which implies that the membrane tension goes to zero, see eq.~\alfa ),\foot{
For simplicity in the presentation, 
we omit the fermion contributions 
(their inclusion is straightforward;  see \refs{\dufi , \russo }).
For further technical details (which are irrelevant for
the scope of this note, such as the gauge fixing of the symmetry of area-preserving diffeomorphisms) we refer to \russo  .
}
\eqn\mmaa{
H_{\rm (A)}= \ha \sum _\n \big( \a^a_{-\n} \a^a_{\n} + \td \a^a_{-\n} \td \a^a_{\n}\big)\ ,\ \ \ \ \ \n\equiv (k,m)\ ,\ \   a=1,...,9\ ,
}
$$
N_\s^+ -N_\s^-= w_0 l_1\ ,\ \ \ \ \ N_\rho^+- N_\rho ^-=  l_2 \ ,
$$
\eqn\zzxx{
N^+_\s = \sum _{m=-\infty }^\infty \sum _{k=1}^\infty {k\ov \ww }
\a^a_\mn \a^a_{ {(k,m)} }\ , \ \ \ \  N^-_\s = \sum _{m=-\infty }^\infty \sum _{k=1}^\infty {k\ov \ww }
\td \a^a_\mn \td \a^a_{ {(k,m)} }\ ,
}
\eqn\ttxxz{
N^+_\rho=\sum _{m=1}^\infty \sum _{k=0}^\infty {m\ov \ww }
\big[ \a^a_\mn \a^a_{ {(k,m)} } + \td \a^a_{(-k,m)} \td \a^a_{(k,-m)} \big]\ ,
}
\eqn\ttxz{
N^-_\rho=\sum _{m=1}^\infty \sum _{k=0}^\infty {m\ov \ww }
\big[ \a^a_ {(-k,m)}\a^a_{(k,-m)} + \td \a^a_{(-k,-m)} 
\td \a^a_{(k,m)} \big]\ .
}
The mode operators satisfy

\eqn\crul{
[ \a _{ {(k,m)} }^a , \a^b_{(k',m')}]= \epsilon (k ) \ww \delta _{k+k'}\delta _{m+m'}\delta^{ab}
\ ,\ }
\eqn\frequ{
\ww =\sqrt { k^2 +  m^2  w_{0}^2 \tau_2^2 }\ ,\ \ \ \ \ \ 
}
where $\epsilon (k )$ is the sign function,  and similar relations 
hold  for the $\tilde \a _{ {(k,m)} } ^a$.

%%%%%%%%%%%%%%%%%%%%%%%%%%%%%%%%%%%%%%%%%%%%%%%%%%%%%%%%%%%%%%
%\newsec{The mass spectrum for the representation (B)}
%%%%%%%%%%%%%%%%%%%%%%%%%%%%%%%%%%%%%%%%%%%%%%%%%%%%%%%%%%%%%%

We  now  calculate the spectrum for the representation (B).
An unboosted  $(q_1,q_2)$ string bound state is now represented
by a static 2-brane with one leg wrapped
around the coordinate $y_1$, and another wrapped around a $(q_1,q_2)$
cycle of the 2-torus
generated by $y_2, y_3$. That is:
$$
y_1(\s + 2\pi ,\rho )= y_1(\s ,\rho )+ 2\pi n R_1'\ ,\ \ $$
$$
y_2 (\s, \rho+2\pi )=y_2 (\s, \rho )+  2\pi q_1 R_2 \ , 
$$
$$
y_3 (\s, \rho+2\pi )=y_3 (\s, \rho )+  2\pi q_2 R_1 \ . \
$$
Adding the boost $w_0$ to the  $(q_1,q_2)$ string bound state
amounts to boosting along the coordinate $y_1 $ with momentum
$w_0/R_1'$. The target 3-torus coordinates can be expanded
as follows:
$$
y_1(\s, \rho )= n R_1' \s + \td y_1 (\s, \rho )\ ,
$$
$$
y_2(\s, \rho )= q_1 R_2 \rho + \td y_2 (\s, \rho )\ ,
$$
$$
y_3(\s, \rho )= q_2 R_1 \rho + \td y_3 (\s, \rho )\ ,
$$
$$
p_{y_1}=\int d\s d\rho \ P_{y_1} ={w_0\ov R_1'}\ ,\ \ \ \ \ 
p_{y_{2,3}}=\int d\s d\rho \ P_{y_{2,3}} = 0\ ,
$$
where $\td y_1, \ \td y_2,\ \td y_3$ are single-valued functions of $\s, \rho $.
Inserting this into the Hamiltonian,
\eqn\lcg{
H =  
2\pi ^2 \int d  \sigma d\rho \left [
 P _ a   ^ 2 + \ha 
{ T_3^2 }
( \{ X ^ a, X ^ b \} ) ^ 2\right ] \ ,
}
and expanding all single-valued functions in $e^{ik\s +im\rho }$,
we now obtain 
\eqn\hbbb{
\a'  H={ (R_1')^2 \ov 2\a '}(l_1^2 +l_2^2\tau_2^2 )+
{1\ov 2} \sum _{\n } \big[ P_{ \n }^a P^a_{-\n }
+\td \ww ^2 X^a_{ \n } X^a_{-\n} \big]+O\bigg({1\ov g}\bigg)
\ , \ \ \ \ 
}
with
\eqn\tdomm{
\td \omega _{km} = \sqrt{
k^2\bigg( q_1^2+q_2^2 {R_1^2\ov R_2^2} \bigg) +m^2 n^2{ R_1 ^{\prime 2}
\ov R_2^2} }\ .
}
In the $g^2\to \infty $ limit, with  $\a ' $ and $R_1'/R_2$  fixed,
the (mass)$^2$ operator $M^2=2H -p_i^2$ then takes the form
\eqn\mmdu{
M^2_{\rm (B)}=M_0^2 + {2\ov \a'} {  H_{\rm (B)}}\ ,
}
\eqn\dmmd{
H_{\rm (B)}= \ha \sum _\n \big( \b^a_{-\n} \b^a_{\n} + \td \b^a_{-\n} \td \b^a_{\n}\big)\ ,
}
\eqn\crul{
[ \b _{ {(k,m)} }^a , \b^b_{(k',m')}]= \epsilon (k ) \td \ww \delta _{k+k'}\delta _{m+m'}\delta^{ab}\ , 
}
with
\eqn\vira{
N_\s^+ -N_\s^-= w_0n\ ,\ \ \ \ \ N_\rho^+= N_\rho ^-\ \ .
}

%%%%%%%%%%%%%%%%%%%%%%%%%%%%%%%%%%%%%%%%%%%%%%%%%
%\newsec{Test of T-duality in the BPS sector}
%%%%%%%%%%%%%%%%%%%%%%%%%%%%%%%%%%%%%%%%%%%%%%%%%

Let us now perform a test of T-duality in the BPS sector.
For a BPS state,  $N_\rho ^-= N_\s^-=0$ 
(which gives the minimum mass for given charges)
and hence  $N_\rho^+= 0$, implying that
such states do not contain excitations $\a _{(k,m)}^a$ with $m\neq 0$.
Thus $\omega _{km}^2 = k^2\big( q_1^2+q_2^2 \tau^2_2 \big)$, or
\eqn\bhuu{
 H_{\rm (B)}=\sqrt{  q_1^2+q_2^2 \tau^2_2 } N_\s^+= w_0 
\sqrt{  l_1^2+l_2^2 \tau^2_2 }\ .
}
Substituting $R_1'=\alpha'/R_1 $ and eq.\bhuu\ into
the mass operator \mmdu , we obtain
\eqn\enne{
M^2_{\rm (B)}= {l_1^2\ov R_1^2} + {l_2^2\ov R_2^2} + {w_0^2 R_1^2\ov 
\a ^{\prime 2}}
+{2\ov \a '}  w_0 
\sqrt{  l_1^2+l_2^2 \tau^2_2 }=\bigg( \sqrt{ {l_1^2\ov R_1^2}+{l_2^2\ov R_2^2} }
+{w_0 R_1\ov \a '} \bigg)^2
\ ,}
which is in striking agreement with the   BPS spectrum of
the $(q_1,q_2)$ string bound states, and thus in agreement
with the BPS oscillation spectrum of the background (A),
as calculated in \rutse .
The matching with the  BPS sector of the spectrum (A)
is not a surprise:  in certain corners
of the moduli parameters ($R_2\to 0$) we must recover exact 
T-duality of perturbative string theory. The BPS mass formula is exact and
should not receive additional corrections as we vary the radius.
Finding different BPS spectra would have implied an inconsistency.
Nevertheless, it is  remarkable that the correct BPS spectrum 
of the $(q_1,q_2)$ string bound state 
(a highly non-perturbative object!) can be derived from 
fundamental supermembranes, and in two inequivalent ways.

%%%%%%%%%%%%%%%%%%%%%%%%%%%%%%%%%%%%%%%%%%%%%%%%%%%%%%%%%%%%%%%%%%%
%\newsec{A discrete spectrum for the thin torus limit}
%%%%%%%%%%%%%%%%%%%%%%%%%%%%%%%%%%%%%%%%%%%%%%%%%%%%%%%%%%%%%%%%%%%
\def\mbb{M_{\rm (B)}}
\def\maa{M_{\rm (A)}}

\medskip

The mass formulas for $\maa $ and $\mbb $
cannot be used to test T-duality in the non-BPS sectors, because
they apply in different corners of the torus moduli parameters.
Although in both cases the relevant limit involves 
the strong coupling limit $g^2\to \infty $ with fixed $\a' $,
for (A) we have kept fixed $R_1/R_2$ (so that $R_1\to \infty $),
while for  (B)  we have fixed $R'_1/R_1=\a'/(R_1R_2)$ (so that
$R_1\to 0$).
If, instead, we take for (B) the limit at fixed $\tau_2=R_1/R_2$,
then flat directions remain in the potential and one obtains
a continuum spectrum in this representation.
Indeed, in this limit
\eqn\contb{
\td \omega_{km}=\sqrt {k^2 \big( q_1^2+q_2^2 \tau_2^2 \big) +         {m^2n^2\over \tau_2^2 g^4} } \ \la \ k \sqrt { q_1^2+q_2^2 \tau_2^2 } \ .
}
Instabilities are then produced by wave packets
constructed with the $X_{(0,m)}^a$  \russo .
Along these directions the harmonic and the quartic terms 
in the Hamiltonian
vanish, and the wave packet can escape to infinity, leading to a continuum
spectrum of eigenvalues. In this $R_1'\to 0$ limit, the method  of small 
oscillations around a stable configuration seems to break down for 
the 2-brane (B).\foot{ This could be related to the  non-renormalizability of
supermembrane theory.}
In the dual description (A), where the membrane is wrapped
around a large area torus, there is nothing pathological and one
obtains an exact discrete spectrum.
If T-duality is a symmetry of M-theory, the mass spectrum of
quantum states associated with background (B) at $R_1'\to 0$  
 must coincide  with that of representation (A).

Similarly, 
in the strong coupling limit at fixed $R'_1/R_2$,
the mass spectrum of representation (A) becomes continuum, because
\eqn\conta{
\ww =\sqrt { k^2 +  m^2  w_{0}^2 \tau_2^2 }\ \la\ k\ .
}
whereas in the same limit
one has an exact discrete spectrum in representation (B).

It is interesting to note that, in the representation (A), we cannot
calculate the spectrum for a membrane with $w_0=0$, because
of membrane instabilities. In the representation (B),
it is possible to study the strong coupling limit
of the excitations of the $(q_1,q_2)$ bound state  with zero momentum $w_0$
from the standpoint of membranes; now the membrane is stable
as long as $l_1\neq 0$ or $l_2\neq 0$, since in this case
the 2-brane winding is non-zero
(it is worth noting that the solution (A) with $w_0=0$
simply represents a gravitational wave;
if T-duality is to hold, the quantum states associated with this background 
in the limit $g^2\to\infty $,
$R_1\to 0$, can also be described  in terms of membrane excitations!).
Although this circumvents the instability problem for this 
sector of the theory, 
the problem subsists for those states with
$l_1=l_2=w_0=0$.

Let us summarize the results and give
 the spectrum in the different corners of
the moduli parameters.
For the sake of clarity, in what follows we consider the simpler
case $l_2=0$.
\medskip

\noindent 1) $R_1, R_2\to \infty $, $T_3\to 0$
($g^2\to \infty $, with $\tau_2=e^{-\phi_0}=R_1/R_2$ and $\a' $ finite): 
\eqn\repra{
\maa ^2=M_0^2 + {1\ov \a' }
\sum _\n \big( \a^a_{-\n} \a^a_{\n} + \td \a^a_{-\n} \td \a^a_{\n}\big)\ ,\
\ \ \ww =\sqrt { k^2 +  m^2  w_{0}^2 \tau_2^2 }\ , }
$$
N_\s^+ -N_\s^-= w_0n\ ,\ \ \ \ \ N_\rho^+= N_\rho ^-\ \ ,
$$
$$
\mbb ^2={\rm continuum \ .}\ \ \ \ \ \ \ \ \ \ \ \ \ \ \ \ \ \ \ \
$$
\medskip
\noindent 2) $R_2\to \infty $, $R_1\to 0$, $T_3\to 0$ 
($g^2\to \infty $, with $R_1'/R_2$ and $\a' $ finite)
$$
\maa ^2={\rm continuum \ ,}\ \ \ \ \ \ \ \ \ \ \ \ \ \ \ \ \ \ \ \
$$
\eqn\reee{
\mbb ^2= M^2_0+ {1\ov \a' }
\sum _\n \big( \b^a_{-\n} \b^a_{\n} + \td \b^a_{-\n} \td \b^a_{\n}\big)\ ,
\ \ \ \td \ww =\sqrt { k^2 +  m^2  n^2 { R_1 ^{\prime 2} \ov R_2^2} } \ ,
}
$$
N_\s^+ -N_\s^-= w_0n\ ,\ \ \ \ \ N_\rho^+= N_\rho ^-\ \ .
$$
\medskip

\noindent 3) $R_2\to 0 $, $R_1\to \infty $, $T_3\to \infty$ 
($g^2 \to 0$,  $\a' $ finite; this is the ten-dimensional limit)
\eqn\nsns{
M^2=M_0^2+  {2\ov \a' } (N_R +N_L)\ ,\ \ \ \ N_R-N_L=nw_0 \ .
}

\medskip
There are  other possible ten-dimensional limits, giving
a similar formula as eq. \nsns .
For example, 
in the limit  $R_1'\to 0$,  $T_3\to \infty $, with 
$\td \a ' \equiv (4\pi ^2 R_1' T_3)^{-1}$  fixed (note that 
$\td \a '=R_1R_2$),
 the relevant degrees of freedom of the system (B)
can be more adequately described by dimensionally reducing
on $y_1$, rather than on $y_2$.
In this process, we find
%$$2+\upa \ \mapa{\rm red} \ 1_{NS}+0_R \ \mapa{T_{z_2}}\ \upa + 1_R  $$
\eqn\araca{
ds^2_{10A}=H_2^{1/2}\big[\td H_2^{-1} \big(-H_2^{-1}dt^2+
dy_2^2\big) + dy_3^2 + dx_idx_i\big]\ ,
}
$$
A=(H_2-1) H_2^{-1}dt\ ,\ \ \ \  B_{ty_2}=\td H_2^{-1}\ ,\ \ \ \ 
e^{2\phi }=\td H_2^{-1} H_2^{3/2}\ ,
$$
which is a bound state of a D0-brane and a fundamental string.
The type IIA string coupling is $\td g^2=R_1'^2/\td \a ' $. 
A T-duality transformation along $y_2$ converts it into a R-R string with charge $w_0$
and a momentum boost $n/R_2'$, with $R_2' = \td\a '/R_2 $.
According to \john , the mass spectrum is then
\eqn\joj{
M^2_{\rm RR}={n^2\ov R_2^{\prime 2}} + 4\pi^2 T^2_{(0,1)} {w_0 R_2^{\prime 2}} +
4\pi T_{(0,1)} (N_R+N_L)\ ,\ \ \ \ N_R-N_L=nw_0
\ ,
}
with $T_{(0,1)}=(2\pi \td \a' )^{-1}{R_2\ov R_1'} $.
In terms of $\a '$, $R_1$, this becomes
\eqn\jojj{
M^2_{\rm RR}=  {n^2\ov R_1^2} + {w_0^2R_1^2\ov \a ^{\prime 2} }  
 +  {2\ov \a' } (N_R +N_L)\ ,
}
i.e. identical to eq. \nsns .

\bigskip

Requiring  exact T-duality symmetry in M-theory
implies  stringent conditions on the underlying fundamental degrees of freedom.
We have  seen that  the mass spectra for the  BPS excitations of 
fundamental 2-branes connected by the T-duality transformations of  eq.~\seis\
 are  T-duality symmetric. 
In addition, T-duality  seems to dictate the precise   
discrete mass spectrum in a regime
where supermembrane theory is not suitable.
This may give a concrete hint on 
an alternative  quantum theory.

%If T-duality symmetry is a correct requirement for M-theory, it 
%implies stringent conditions on the underlying fundamental degrees of freedom.

\bigskip

%%%%%%%%%%%%%%%%%%%%%%%%%%%%%%%%
\noindent $\underline {\rm Acknowledgements}$: 
%%%%%%%%%%%%%%%%%%%%%%%%%%%%%
I would like to thank K. Stelle and A. Tseytlin for useful discussions.

\listrefs

\bye